\documentclass[12pt,fleqn]{article}

\usepackage{amsmath}
\usepackage{amssymb}
\usepackage{amsfonts}
\usepackage{afterpage}
\usepackage{graphicx}

\newcommand{\PSImagx}[2]{\includegraphics[width=#2]{#1.ps}}

\newcommand{\BILD}[4]{\begin{figure}[#1]%

          #2

          \centerline{\parbox{16cm}{\caption[.]{#3} \label{#4}}}
          \end{figure} }

\newcommand{\X}{{X}}

\newcommand{\TrM}{{\rm Tr }M}

\newcommand{\Wpn}{W^{\text{p}}_n}
\newcommand{\Wfn}{W^{\text{f}}_n}
\newcommand{\Wf}{W^{\text{f}}}

\newcommand{\diag}{\mathop{\rm diag}}
\newcommand{\param}{\varepsilon}
\newcommand{\limacon}{lima\c{c}on }

\begin{document}
\noindent


~ 

\vspace*{1cm}

\newcommand{\titel}{About ergodicity in the \\[0.5ex]
                         family of \limacon billiards}


\normalsize

\vspace{0.5cm}

\renewcommand{\thefootnote}{\fnsymbol{footnote}}

\begin{center}  \huge  \bf

           \titel
\end{center}

\setcounter{footnote}{1}

\begin{center}
        \vspace{2cm}

              { \large Holger R.~Dullin%
\footnote[1]{E-mail address: {\tt H.R.Dullin@lboro.ac.uk}}
               }
              {\large and Arnd B\"acker%
\footnote[7]{E-mail address: {\tt a.backer@bristol.ac.uk}}$^{,\ast\ast\ast}$
              }

        \vspace{6ex}

        $^{\ast}$ Department of Mathematical Sciences, Loughborough University\\
        Loughborough, Leics, LE11 3TU, UK \\
        \vspace{4ex}

        $^{\ast\ast}$
     School of Mathematics, University of Bristol\\
     University Walk, Bristol BS8 1TW, UK\\
        \vspace{4ex}

     $^{\ast\ast\ast}$ BRIMS, Hewlett-Packard Laboratories\\
     Filton Road, Bristol BS12 6QZ UK\\

\end{center}

\vspace*{3cm}

\leftline{\bf Abstract:}

By continuation from the hyperbolic limit of the cardioid billiard
we show that there is an abundance of bifurcations
in the family of \limacon billiards.
The statistics of these bifurcation shows that
the size of the stable intervals decreases with approximately the same rate
as their number increases with the period.
In particular, we give numerical evidence that arbitrarily close
to the cardioid there are elliptic islands due to orbits created
in saddle node bifurcations.
This shows explicitly that if in this one parameter
family of maps ergodicity occurs for more than one parameter
the set of these parameter values has a complicated structure.

\renewcommand{\thefootnote}{\arabic{footnote}}
\setcounter{footnote}{0}


\newpage

\section{Introduction}

In Hamiltonian systems ergodicity is a rather rare property \cite{MarMey74}.
Most systems are ``mixed'',
in the sense that there are regions of regular motion
of positive measure and regions with stochastic motion,
which are expected to be of positive measure, too;
see \cite{Str91} for a review on the coexistence problem.

In this paper we investigate the possibility of ergodicity in the family
of billiards introduced by Robnik \cite{Rob83}, which is
given by the simplest non-trivial conformal mapping
of the unit circle; in polar coordinates
the boundary of the  family of \limacon billiards is parametrised by
\begin{equation} \label{eqn:rhophi}
       \rho(\phi) = 1 +  \param \cos (\phi) \;\;, \qquad \phi 
\in[-\pi,\pi] \;\;,
\end{equation}
where $\param\in[0,1]$ is the family parameter.
This family of billiards has received a lot of attention, especially in
the context of quantum chaos, see e.g.\
\cite{Rob84,BruWhe96,Bae98:PhD},
and the references therein.
Corresponding to $\param = 0$ is the integrable billiard in the circle,
while $\param=1$ gives the ergodic billiard inside the cardioid
\cite{Woj86,Sza92,Mar93}.
For a small perturbation of the integrable case the KAM theorem
implies that most invariant tori persist,
whereas for the neighborhood of the hyperbolic case there is no
general result about persistence of non-uniformly hyperbolicity.
    From this point of view one of the most interesting questions
for this billiard family is whether it
is hyperbolic or even ergodic for some $\param <1$.

For $\param\in[0,\frac{1}{2}[$ the boundary of the billiards is convex,
$C^{\infty}$--smooth and the curvature is strictly positive.
As a consequence of a general theorem by Lazutkin \cite{Laz73a}
these billiards have a collection of caustics
near the boundary, which correspond to rotational invariant circles
in the Poincar\'e map from bounce to bounce.
Since rotational invariant circles divide the Poincar\'e section into
invariant regions these billiards are not ergodic.
At $\varepsilon=\tfrac{1}{2}$ the boundary has zero curvature at
$\phi=\pm \pi$ and therefore the invariant circles corresponding
to the caustics near the boundary no
longer persist \cite{Mat82}.
For $\varepsilon>\tfrac{1}{2}$ the billiards are no longer convex,
and the billiard map becomes discontinuous.
     From these observations and from numerical experiments
Robnik suggested that for $\varepsilon>\tfrac{1}{2}$ the
billiards might be ergodic \cite{Rob83}, see also \cite{ProRob94b,LiRob95}.
It turned out, however, that at least close to
$\varepsilon\ge\frac{1}{2}$ this is not true
because still a family of Liapunov stable periodic orbits exists
\cite{HayDumMouStr87}. This orbit
is involved in a cascade of period doubling bifurcations, where the
extrapolated limit is $\varepsilon=0.5582\ldots$ \cite{HayDumMouStr87}.
A first example of a stable periodic orbit above this value
was given in \cite{KlaSmi96} where a bifurcation is discussed
for $\varepsilon\approx0.66948$.

One way to prove that the system is not ergodic is to establish
the existence of an invariant curve.
This is possible by using KAM theory in the neighborhood of an elliptic
fixed point of the map. A sufficient condition for the existence
of an invariant curve in the neighborhood of an elliptic fixed point
is that the eigenvalue of the linearised map at the fixed point
is not a third or forth root of unity {\em and\/} that the twist does
not vanish, see \cite{SieMos71}.
In \cite{HayDumMouStr87} this was rigorously established
for a certain orbit of the \limacon family.
In the following we will rely on numerical results in which we
cannot easily check the twist condition.
However, as in one parameter families the twist generically vanishes at points
\cite{DulMeiSte2000},
the existence of an elliptic orbit for one given
parameter value gives
strong numerical evidence that the map is not ergodic
at this parameter.

As for $\varepsilon$ larger than the extrapolated limit of the
period--doubling cascade there are no obvious stable islands in the Poincar\'e
section, it might be conjectured that at this point
all the orbits that are part of the bifurcation tree
that is connected to the integrable circle billiard have turned unstable.
But even if this was true (for which we have not found any counterexamples),
it does not imply ergodicity of the system. The reason is that
there exist stable orbits which are not connected to the integrable limit.
They appear in saddle--node bifurcations, in which a stable and unstable
orbit are created  ``out of nowhere'', in principle at any parameter value.
At each saddle--node bifurcation the newly created elliptic orbit
exists in a certain parameter interval until it turns unstable in
a period doubling bifurcation.
We give numerical evidence that there are
saddle--node bifurcations arbitrarily close to the ergodic
limit at $\param  = 1$. Moreover, our numerical
computations show that the sum of the widths
of the stable intervals of elliptic orbits born in
saddle--node bifurcations has considerable measure in parameter space.
For a related study of the phenomenon of the appearance of elliptic islands
in systems close to ergodic billiards see \cite{TurRom98} or in
the standard map with large parameter see \cite{Dua94}.

However, we cannot rule out that ergodicity holds for some values
$\param < 1$, because the stable intervals of these orbits are so small
that they do not seem to cover the whole interval $[\tfrac{1}{2},1[$.
The method we use is
continuation of  periodic  orbits away from the hyperbolic limit.
At this limit we know all periodic orbits.
For the regular orbits, i.e.\ those orbits which never hit
the cusp of the cardioid,
a binary symbolic dynamics \cite{BruWhe96,BaeDul97}
was proven in \cite{BaeChe98,Dul98b}, and for the cusp orbits
a similar coding with an additional letter can be used.
The idea to continue periodic orbits from hyperbolic a hyperbolic
limit is similar to continuation from an antiintegrable limit \cite{Aub95},
which has recently be applied to the H\'enon map \cite{SteDulMei99}.
However  for an unbounded system like the H\'enon map,
the map can actually be proven to be hyperbolic near its limit
\cite{DevNit79,SteMei98}.
For a system with bounded phase space this is rather unlikely due to the
conservative Newhouse phenomenon \cite{Dua99}.
Moreover, the cardioid billiard is {\em not} uniformly hyperbolic.
There is no positive lower bound for the Liapunov exponent of certain orbits.
Nevertheless, for the cardioid every finite
periodic orbit is hyperbolic and therefore can be continued.
    But due to the non-uniform hyperbolicity, the
system can become non-hyperbolic when perturbed.

\section{Finding and continuing orbits}

Our approach is to continue periodic orbits of
the cardioid billiard ($\param=1$) to smaller values of the parameter $\param$.
For the continuation we use the discrete variational principle, which is
numerically much more stable than iterating the map, and also allows for the
continuation of singular orbits. The discrete analogue of Hamilton's principle
states that the action
\begin{equation}
      W =  \sum_{i = t_1}^{t_2}  L(x_i,x_{i+1}) \;\;,
\end{equation}
is stationary on orbits, where $L$ is the discrete Lagrangian of the system.
Performing the variation $\delta W = 0$ gives a second order
difference equations.
Assuming the twist condition holds the analogue of the Legendre transformation
can be used to rewrite it as an area preserving map, see e.g.\
\cite{Mei92} for a review.

For billiards the discrete Lagrangian $L(x,x')$ is just the geometric
length between the two points on the boundary parameterised by $x$ and $x'$,
so that $W$ is the total length of the orbit.
The Poincar\'e map from bounce to bounce of a smooth billiard
is a map of a finite cylinder to itself.
      The global winding number of a periodic orbit is given by $\frac{m}{n}$,
      where $n$ is the number of reflections and $m$
      counts how often the orbit goes around the cylinder.

We use polar coordinates $(\rho,\phi)$ to describe the billiard boundary,
so that the polar angle  $\phi$ replaces  $x$.
With (\ref{eqn:rhophi}) the discrete Lagrangian becomes
\begin{equation}
	L(\phi,\phi') = \sqrt{\rho(\phi)^2 + \rho(\phi')^2 -
		2\rho(\phi)\rho(\phi')\cos(\phi'-\phi)} \;\;.
\end{equation}
For a doubly periodic Lagrangian periodic orbits of period $n$
are critical points of the periodic action
\begin{equation}
       \Wpn({\X_n})  = L(x_0,x_1) + L(x_1,x_2) + \dots + L(x_{n-1}, x_0 )
            \;\;,
\end{equation}
where $\X_n = (x_0, x_1, \dots, x_{n-1} )$.
In general the critical points of $\Wpn$ are found using Newton's
method applied to the system of equations $\nabla \Wpn = 0$.
In \cite{BaeDul97,Dul98b} it was shown that all regular orbits,
i.e.\ those which never hit the cusp, of the cardioid
are {\em maxima} of $\Wpn$, such that much more stable numerical
methods can be used to find them.

Let us now briefly recall the definition of the symbolic dynamics
for the cardioid billiard \cite{BruWhe96,BaeDul97}
which gives a complete description of all possible orbits in the cardioid,
see \cite{BaeChe98,Dul98b} for proofs.
Consider two consecutive points on the boundary with polar angles
$\phi,\phi' \in (-\pi,\pi]$.
If $\phi = \pm\pi$ or $\phi'=\pm\pi$ the symbol C is associated.
Otherwise the corresponding symbol is A if $\phi'<\phi$ and B if $\phi'>\phi$.
The method described in \cite{Dul98b} requires to find initial conditions
consistent with a given code word, for which we use the following procedure.
Rotate the string until it starts with BA.
In step 1 choose $(\phi_0,\phi_1) = (z,\delta)$ with $z > \delta > 0$
as the initial pair
corresponding to the letter A. Here $\delta$ is a small positive number
and $z$ will be determined at the end.
In step $i$ take $\phi_{i-1}$ from the previous step, and construct the pair
$(\phi_{i-1},(\phi_{i-1}-\pi)/2)$ for A and
$(\phi_{i-1},(\phi_{i-1}+\pi)/2)$ for B.
In the last step $n$ we obtain by construction the letter B, which is the
first letter of the word. We have to fulfill the conditions
$\phi_{n-1} < z$ and
$z > \delta$, e.g.\ by the choice $z = (\max(\phi_{n-1},\delta) + \pi)/2$.
Note that all $\phi_i$ are in $[-\pi,\pi]$ for arbitrary sequences of A and B.
This shows that we can choose an admissible initial condition for
each code word, i.e.\ there is no intrinsic pruning \cite{Dul98b}.

If the maximization of $W$ for a given symbol sequence leads
to a solution for which one or more segments
run outside of the billiard, this periodic orbit is not realised
by the system; the orbit is called ``f--pruned'' \cite{BaeDul97}.
It may also occur that the solution possesses a point in the singularity,
a situation in which we say the orbit is ``s--pruned'',
or in the more general context of \cite{Dul98b} extrinsically pruned.
Also combined types of pruning occur, see \cite{BaeDul97}
for examples of the different types of pruning.

In order to find all orbits for $\param < 1$ we also have to
find finite orbits in the cardioid billiard, i.e.\ cusp orbits,
which begin and end with an orbit segment in the cusp,
since most of them turn into regular orbits for $\param<1$.
This might seem surprising at first; see below for a proof.
The main difference between these singular orbits and the regular ones
it that there is no reflection condition to be fulfilled for orbit segments
in- and outgoing to the singularity of the cardioid at $\phi_i = \pm \pi$
because
\begin{equation} \label{eqn:norefl}
	\left.
	  \frac{\partial \Wpn}{\partial \phi_i}
\right|_{\phi_i=\pm\pi} = \left.
	  \frac{\partial L(\phi_{i-1},\phi_i)}{\partial \phi_i} +
	  \frac{\partial L(\phi_{i},\phi_{i+1})}{\partial \phi_i}
             \right|_{\phi_i=\pm\pi} = 0 \;\;.
\end{equation}
In order to find cusp orbits we can therefore look for critical points of
an action with fixed initial and final point
\begin{equation}
	\Wfn(\X_n;\alpha,\omega)  =  L(\alpha,x_0) + L(x_0,x_1) + \dots
		+ L(x_{n-1},\omega)
	\;\;.
\end{equation}
Note that $\Wfn$ describes an orbit with $n$ reflections not counting
reflections at $\alpha$ and $\omega$.
It is sufficient to restrict to cusp orbits
hitting the cusp just once,
as these can be combined
to multiple cusp orbits by concatenation.
Similar to the case of regular
orbits in the cardioid we can find cusp orbits by looking for maxima of $\Wfn$.
The proof is as for $\Wpn$,
the only difference is that $L(\phi_{n-1},\phi_0)$
is replaced by $L(\phi_{n-1},\omega) + L(\alpha,\phi_0)$. For the cardioid
with $\alpha = \omega = \pi$ the Hessian of this function is
$\diag(-\cos \phi_0,-\cos \phi_{n-1})$.
Thus it is negative definite
for orbits which hit the cusp from the right
(both angles, before and after hitting the cusp,
have modulus smaller than $\pi/2$) and the same proof as in
\cite{Dul98b} works.
Unfortunately we have not been able to show that orbits
with modulus of $\phi_0$ or $\phi_{n-1}$ larger then $\pi/2$ are also minima.
Even if such an orbit is initially pruned,
e.g.\ $A^{2n}CC$ for $n\ge2$, it can become physical under continuation.
The shortest code for which we cannot
find a corresponding minimum in the action is $A^6B^3CC$. However,
using the symmetry line method described in \cite{Bae98:PhD} to find
this orbit shows that it is in fact s-pruned.

The initial conditions for the numerical determination
of the maximum of $\Wfn$ for a given cusp orbit
are obtained similarly to the case of regular orbits:
first the code word is rotated so that it starts with CC.
The same procedure as above is applied for intermediate $\phi_i$, except
in the end we just put $\phi_0=z=\pi$.
To follow orbits under parameter variation we use a
predictor--corrector method, see e.g.\ \cite{Sey94}.
For a slightly changed parameter $\param+\Delta\param$
an initial guess for the orbit is computed
in the predictor step, which is used in the corrector step
to find the actual orbit at the new parameter.
We continue the orbit $\Phi = (\phi_0,\dots,\phi_{n-1})$
under parameter variation in $\param$ by expanding
\begin{equation} \label{eqn:continue}
	\nabla \Wpn(\Phi + \Delta\Phi; \param + \Delta \param) \approx
	\nabla \Wpn + D^2\Wpn \Delta\Phi +
                \frac{\partial \nabla \Wpn}{\partial \param} \Delta \param \;\;.
\end{equation}
Because $\Phi$ is an orbit at parameter $\param$ we have $\nabla \Wpn = 0$.
Hence we can solve for $\Delta \Phi$  provided that $D^2\Wpn$ is nonsingular.
This gives the predictor $\Phi + \Delta\Phi$ for the continued orbit at
parameter $\param+\Delta \param$.
The corrector step consists in Newton's method
taking the predictor as the
initial guess for the orbit at the new parameter $\param+\Delta \param$.
Note that for continuing orbits away from the cardioid it is
necessary to use Newton's method, because for $\param<1$
not all the critical points of $\Wpn$ are maxima.
If the orbit cannot be found for the new parameter $\param+\Delta \param$
the step size is decreased.
If under parameter variation $\det D^2\Wpn = 0$ is reached, we have
found the point (in most cases, see Sec.~\ref{sec:bifurcations})
at which the orbit is created.
In order to be able to continue orbits through this singular point in the
parametrization  by
family parameter $\param$ we consider the augmented problem,
see e.g.\ \cite[chpt.\ 4.4]{Sey94},
in which the parameter is itself a variable,
and the continuation uses the
arc length along the curve in $(\Phi,\param)$ space as a new parameter.

A cusp orbit of length $n$ has $n-1$ points $\Psi_{n-1} =
(\psi_0,\dots,\psi_{n-2})$
and one point at the cusp. $\Psi_{n-1}$ satisfies the equation
$\nabla \Wf_{n-1}(\Psi_{n-1};\pi,\pi) = 0$ at $\param = 1$.
To continue a cusp orbit away from the cardioid we
make the fixed angle $\pi$ in $\Wf_{n-1}(\Psi_{n-1};\pi,\pi)$
an additional variable and extremise $\Wpn(\Phi_n)$
for $\param < 1$, where initially $\Phi_n = (\Psi_{n-1},\pi)$.
Note that because of (\ref{eqn:norefl}) we have
$\nabla \Wpn(\Phi_n) = 0$ at $\param=1$ as before.
However, the Hessian of $\Wpn$ is not definite at this point.
For the continuation the important fact is that it is nondegenerate,
which we will now show.
First note that
\begin{equation} \label{eqn:nohefl}
	\left.
	  \frac{\partial^2 \Wpn}{\partial \phi_i \partial \phi_j}
             \right|_{\phi_i=\pm\pi}
= \left\{ \begin{array}{ll}
\cos\phi_{i-1}+\cos\phi_{i+1} & \text{if $i=j$} \\
                                0 & \text{else}  \end{array}  \right.
      \, ,
\end{equation}
and that outside the cusp the second
derivatives of $\Wfn$ and $\Wpn$ coincide.
Therefore we find
\[
        D^2 \Wpn|_{\Phi_n} = \begin{pmatrix} \vartheta & 0 \\ 0 & D^2
\Wf_{n-1}|_{\Psi_{n-1}} \end{pmatrix}
\quad \text{where} \quad \vartheta = \cos\psi_0+\cos\psi_{n-2} \,.
\]
We have already shown that $D^2\Wf_{n-1}$ is negative definite, and
we also noted that for physical orbits the cosines in $\vartheta$ are both
positive. Therefore
$D^2\Wpn$ is nondegenerate at cusp orbits. The difference to orbits that do
not hit the cusp is that here the Hessian is not negative definite,
instead it has
one positive eigenvalue. In any case these orbits can be continued
away from the
cardioid. By obvious extension of the above the case where a finite orbit hits
the cusp more than once can be included as well.

It is surprising that the continuation of cusp orbits away from the
cardioid is not singular, because these orbits are ``infinitely
unstable'' in the cardioid.
This can be seen using a formula for the residue $R$ of an orbit
\cite{KayMei83}
\[
         R = -\frac{1}{4} \frac{\det D^2 \Wpn }{ \prod_{i=0}^{n-1} B_i},
\quad \text{where} \quad
      B_i = \frac{\partial^2 \Wpn}{\partial \phi_i \partial \phi_{i+1} }
\]
with cyclic boundary conditions. Observe
that for $\param \to 1$ we have that $B_{0} \to 0,  B_{n-1} \to 0$
due to (\ref{eqn:nohefl}) and thus $R$ diverges.
The limiting behaviour can be found by expansion around
$\param = 1$. This gives $B_i = (1-\param)\cos\phi_i$ for $i=0$ and $i=n-1$,
so that we find
\[
        R \approx -\frac{1}{4(1-\param)^2} \frac{\det D^2 \Wf_{n-1}}{
\prod_{i=1}^{n-2} B_i}
               \left( \frac{1}{\cos\phi_0} + \frac{1}{\cos\phi_{n-1}} \right)
\]
near $\param = 1$. Note that for multiple cusp orbits a similar
result holds with
the power of $1-\param$ decreased by one for each point in the cusp.

\section{Bifurcations of orbits} \label{sec:bifurcations}

\BILD{pq}
          {
            \centerline{\PSImagx{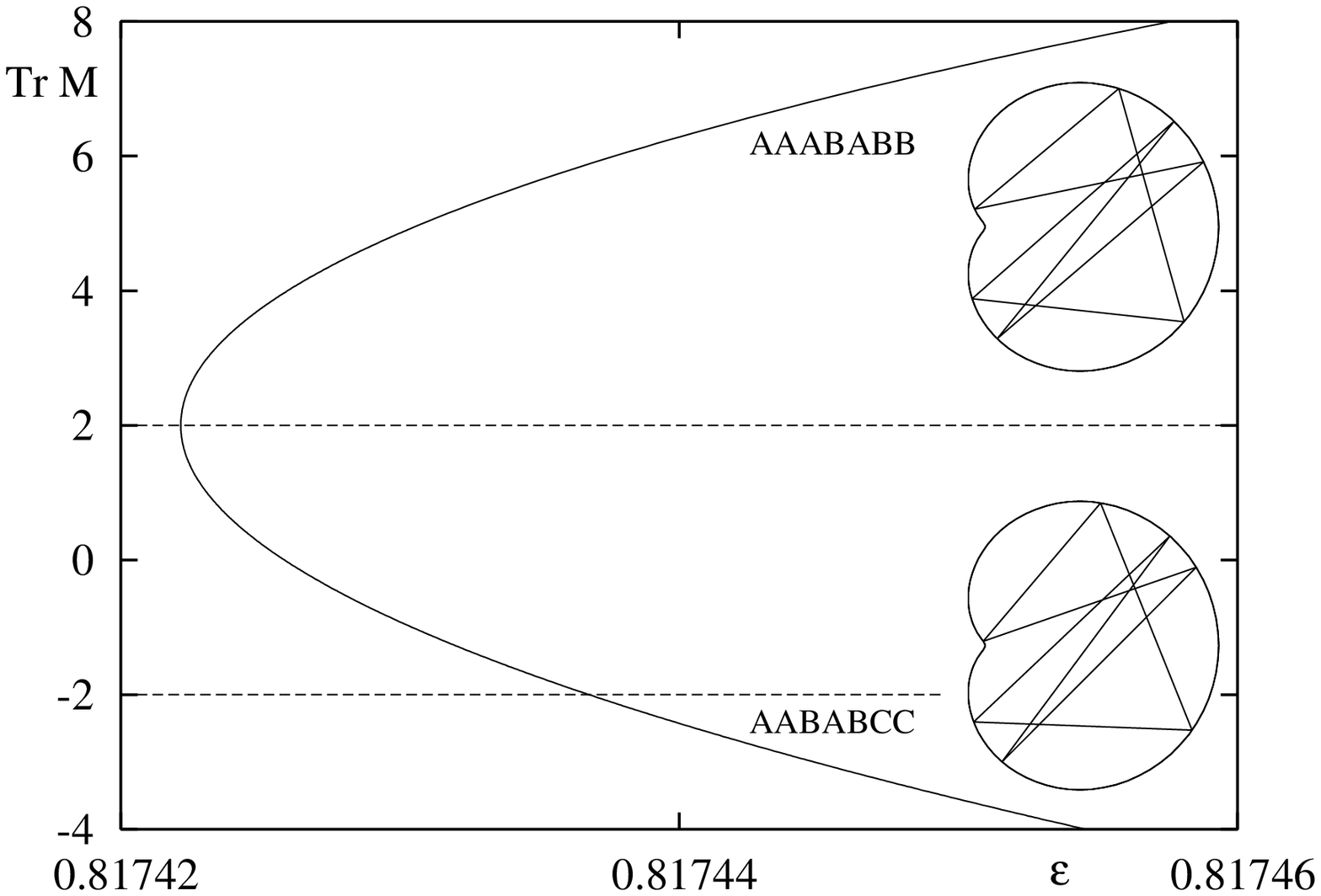}{12cm}}
          }
          {\small The trace $\TrM=-4R+2$ of the monodromy matrix
           is shown for the shortest nonsymmetric periodic orbit $AAABABB$
           and the cusp orbit $AABABCC$ in the cardioid
           under variation of the parameter $\varepsilon$.
           At $\varepsilon=0.817422\ldots$ these periodic orbits are created
           in a saddle--node bifurcation (sn).
           Also shown are the corresponding periodic orbits
           at $\varepsilon=0.85$.
                }
          {fig:bif_diag_AAABABB_AABABCC}

\BILD{p}
          {

            \centerline{\PSImagx{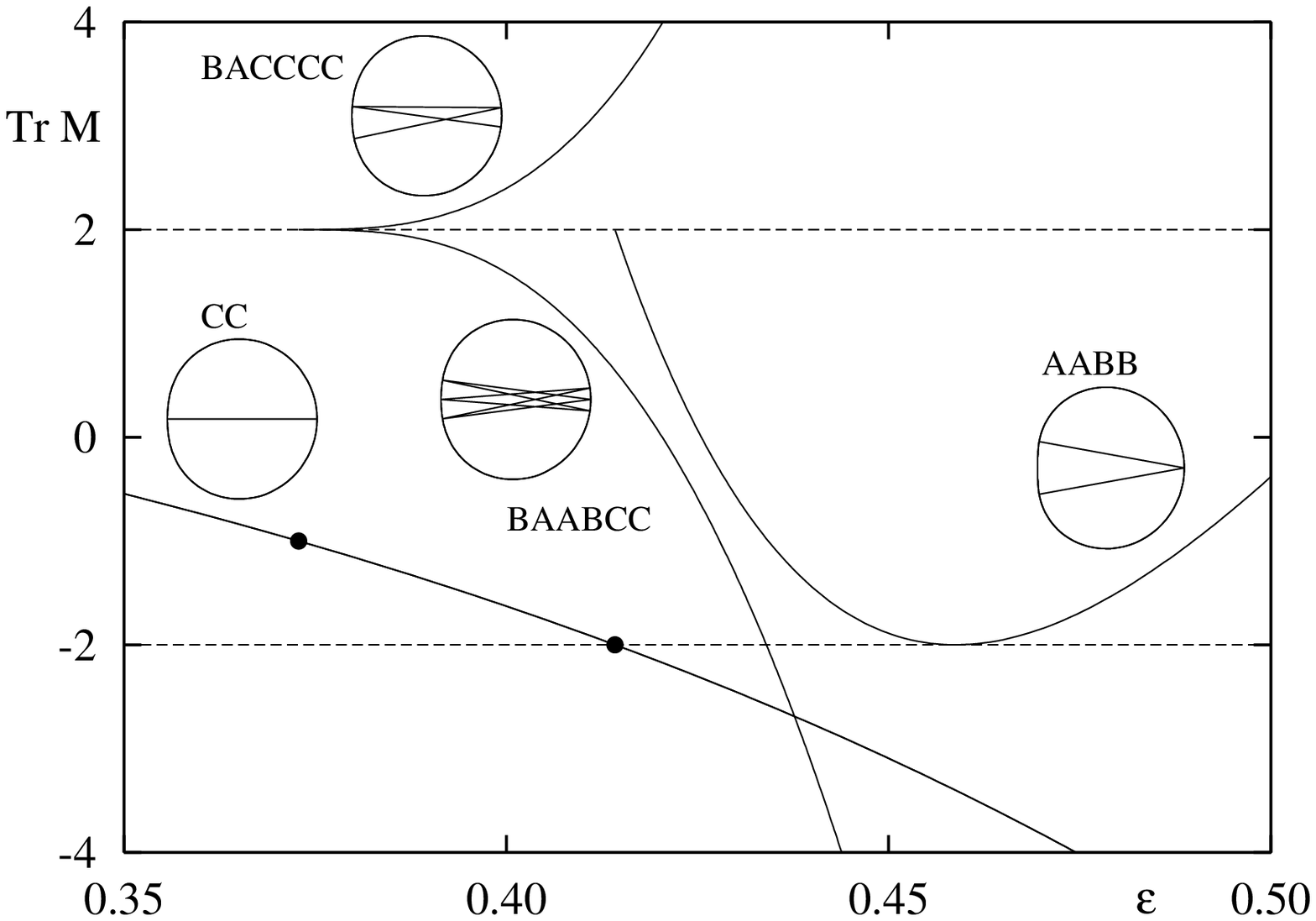}{12cm}}
          }
          {\small Under parameter variation the $AB$ orbit has a
symmetry breaking
            period tripling bifurcation (s1/3) at $\param = 0.3728\dots$
creating the unstable
            orbit $BACCCC$ and the initially elliptic orbit $BAABCC$. At
$\param = \sqrt{2}-1$
            a period doubling (pd) bifurcation gives birth to the $AABB$
orbit. This orbit
            has full symmetry and therefore has a tangency with $\TrM =
-2$ at $\param = 0.4587\dots$.
                }
          {fig:bif_diag_AB}

\BILD{p}
          {
            \centerline{\PSImagx{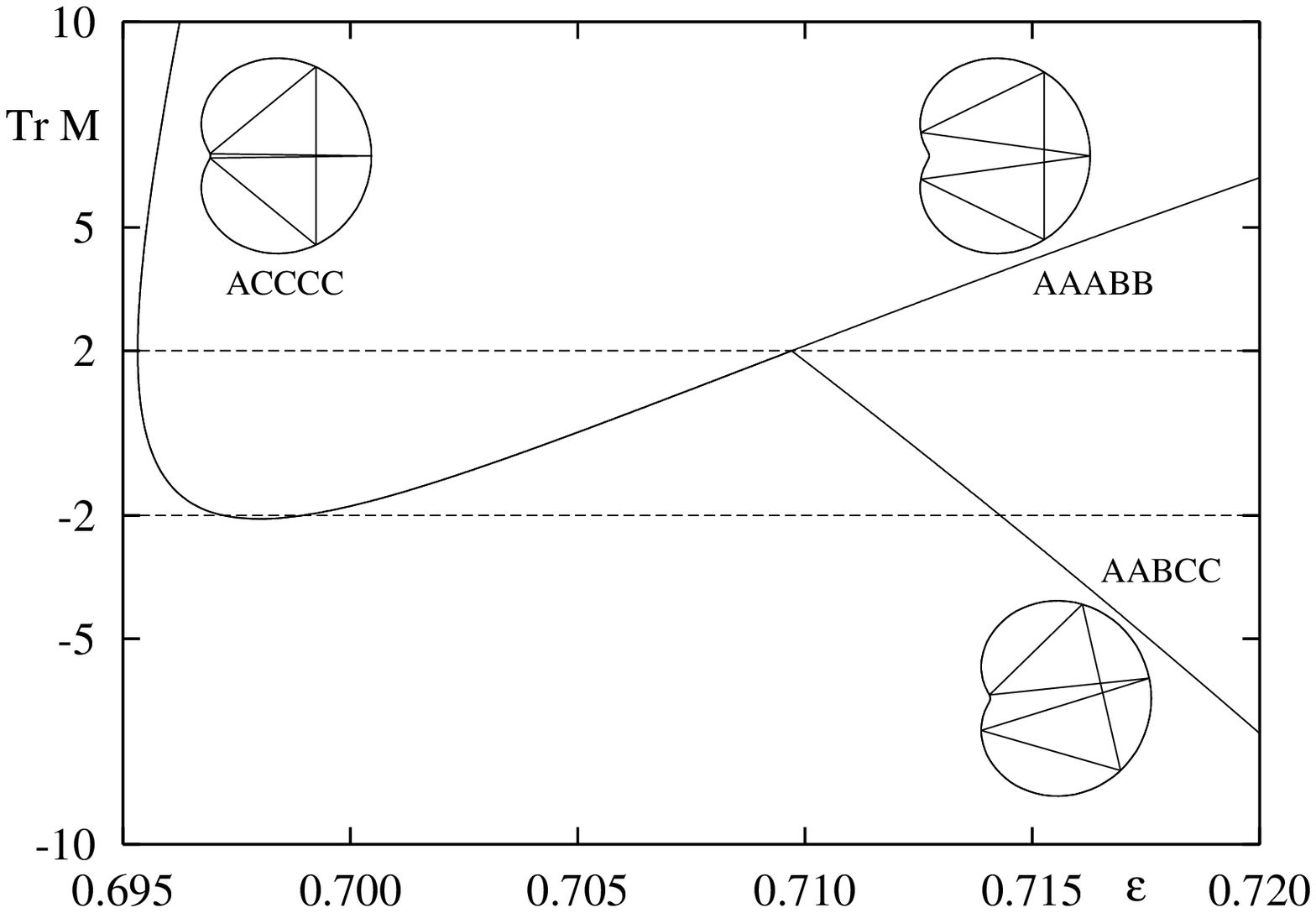}{12cm}}
          }
          {\small Example of another typical bifurcation scenario for
orbits with symmetry.
           First at $\varepsilon=0.6953235\ldots$ the two orbits $AAABB$
(stable) and
           $ACCCC$ (unstable) are created in a saddle--node bifurcation.
           Notice that the orbit $AAABB$ becomes inverse hyperbolic
           but again elliptic at $\param = 0.697988\dots$.
           At $\varepsilon=0.709705\ldots$
           a symmetry breaking pitchfork bifurcation (pf) takes place
           in which the elliptic $AABCC$ (and its symmetric partner
$BAACC$) are created.
           All orbits are shown for $\varepsilon=0.8$.
          }
          {fig:bif_diag_AAABB_ACCCC}

\BILD{p}
          {
            \centerline{\PSImagx{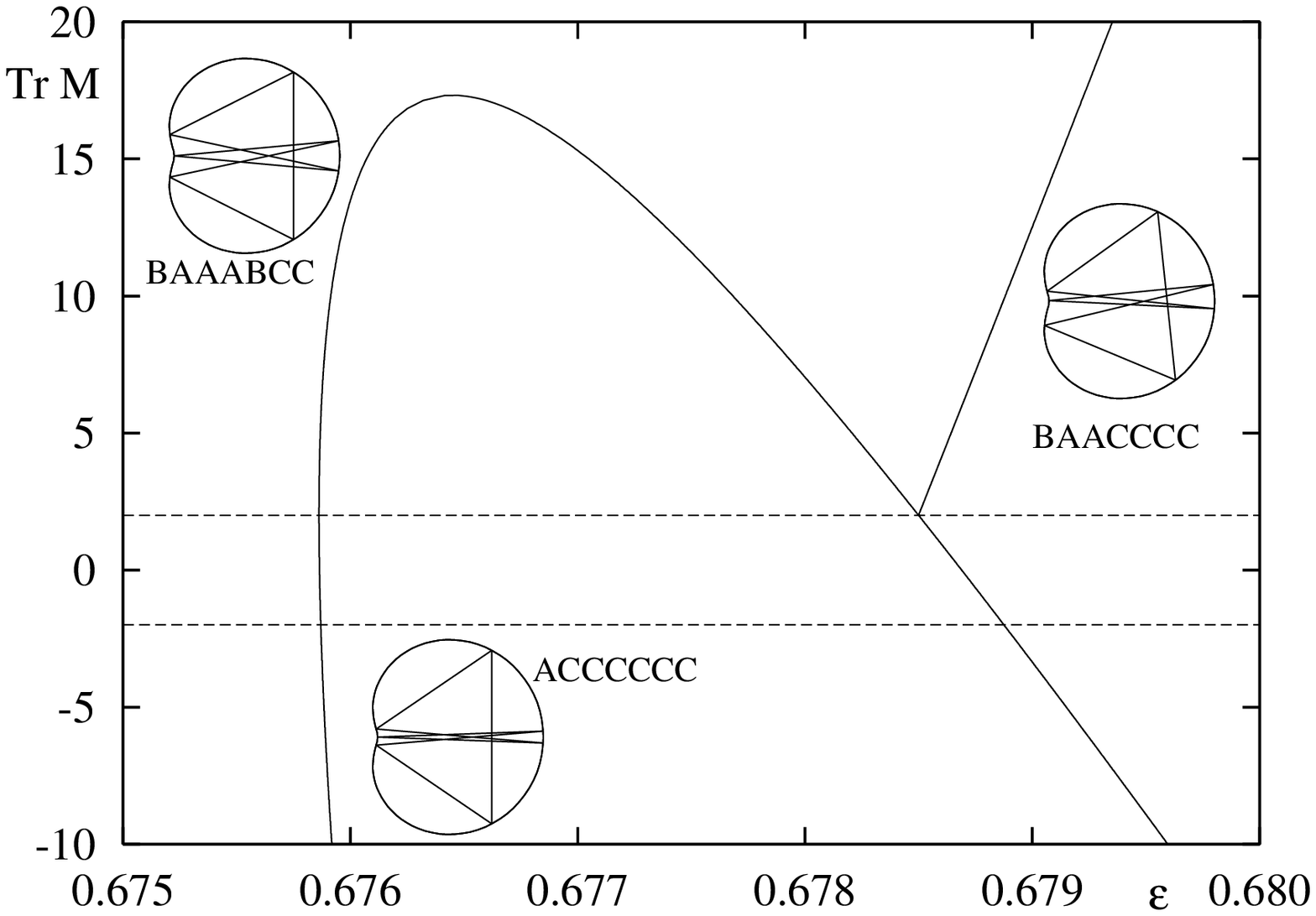}{12cm}}
          }
          {\small The bifurcation diagram for the unstable pitchfork
                bifurcation (upf)
                is the mirror image of the previous one in 
Fig.~\ref{fig:bif_diag_AAABB_ACCCC}.
                Here either partner in the saddle-node bifurcation
                  ($\param = 0.675862\dots$)
                eventually becomes stable.
                The initially unstable orbit $BAAABCC$ becomes stable at
                $\param = 0.678499\dots$
                 and creates $BAACCCC$ and its symmetric partner
                 $AABCCCC$ (not shown).
                }
          {fig:bif:diag:ACCCCCC}

\afterpage{\clearpage}

To study the bifurcations of a periodic orbit
we start in the cardioid and continue it
from $\param=1$ towards the circle $\param=0$ as long as it exists.
We include all cusp orbits and also
all pruned orbits, because they typically become physical
for smaller $\param$ when the billiard becomes closer to being convex.
An orbit can cease to exist if it collides with another one at
residue $R=0$, corresponding to an eigenvalue 1 of the linearised
Poincar\'e map.
For all other finite $R$ orbits
cannot disappear, which can be seen by
applying the implicit function theorem to (\ref{eqn:continue}).
This only holds for smooth maps, so the singular bifurcations
which occur in a non--convex billiard when there
is a tangency of an orbit segment and the boundary
can occur with any residue.
The variational approach is blind to these singular bifurcations,
except if there is a reflection at the point of tangency.
In this case the stability of the orbit diverges
at the tangency.
Numerically the continuation is done for decreasing $\param$.
However, when describing bifurcations we always consider an
increasing parameter.
The following possibilities arise:
\begin{itemize}
\item The orbit exists down to $\param=0$.
           In the integrable circle billiard
           periodic orbits can be labeled by their rational
          	rotation number $\omega = p/q$.
           Under small perturbations the corresponding rational torus breaks
           into a Poincar\'e-Birkhoff chain
           consisting of one unstable and one stable orbit of period $q$.
           The existence of rotational $p/q$ orbits for larger $\param$ is
           guaranteed by Aubry-Mather theory \cite{Aub83,Mat82b}
           (see \cite{Mei92} for a review)
           as long as the billiard
           is strictly convex,
           i.e.\ for $\param<1/2$.
            The unstable $p/q$ orbits maximise the action $W$, the
           (initially, i.e.\ for $\param>0$, $\param$ small)
           stable ones are called minimax, they are saddles of $W$
           with one up direction.
           The minimax orbits up to period 7 together with the
           $\param$ value at which they turn unstable are listed in
           Table~\ref{tab:minimax}.

\item The orbit only exists down to a nonzero parameter value
           $\param_{\text{c}}$
           at which it has $R=0$ and $\partial R/\partial \param \not = 0$.
           In systems with a reversing symmetry generically one of the
following bifurcations
           occurs; the orbit collides with
           \begin{enumerate}
             \item another orbit
                with opposite sign of $R$ in a saddle node bifurcation (sn);
             	  there is no parent orbit (Fig.~1).
             \item another orbit with the same sign of $R$ in a 
symmetry breaking
                pitchfork bifurcation (pf); the parent orbit has $R=0$ (Fig.~3).
                The subcritical pitchfork bifurcation
                in which the parent turns stable is denoted by upf (Fig.~4).
             \item itself in a period doubling bifurcation (pd);
             	     the parent orbit has $R=1$. The subcritical
                   period doubling in
                   which the parent turns stable is denoted by upd (Fig.~2).
             \item another orbit with opposite sign of $R$ in a
                   $p/q$-tupling bifurcation,
           	      $q\ge 3$; the parent orbit has $R=\sin^2(\pi p/q)$.
           The ``touch and go'' bifurcation with
           $\partial R/ \partial \param = 0$
           for $q=3$ and possibly $4$ does not terminate the orbit 
because it also
           exists for smaller $\param$.
           \end{enumerate}
\end{itemize}
The above cases exhaust the list of generic
bifurcations with one symmetry \cite{Rim83}. The family of \limacon billiards
has two reversing symmetries, $X$ and $T$, see \cite{BaeDul97}.
Therefore there are more generic bifurcations of symmetric orbits in
this family
\cite{AguMalBarDav87,AguMal88,The99:Dip}. E.g., the period tripling of
an orbit with
full symmetry creates four orbits with less symmetry whose phase portrait looks
like a six-tupling bifurcation. We denote this bifurcation by s$1/3$ (Fig.~2).

\begin{table}[b]
\centerline{\small
\begin{tabular}{c|c|c|c}
$p/q$ & code & $\param$ & $\param_{\rm tangency}$ \\ \hline
1/2 & CC    & 0.41421356237 & --- \\
1/3 & CAC   & 0.36810610044 & ---\\
1/4 & CAAC  & 0.35203466109 & ---\\
1/5 & CAAAC & 0.34773961433 & ---\\
2/5 & CABAC & 0.30229901387 & ---\\
1/6 & CAAAAC& 0.34830176077 &
                  0.89431500512 \\
1/7 &CAAAAAC& 0.35095913185 &
                  0.81112873450 \\
2/7 &CAABAAC& 0.29265846151 & ---\\
3/7 &CABABAC& 0.27009692367 & ---
\end{tabular}}
\caption{Minimax orbits with winding number $p/q$, the corresponding
code in the
cardioid billiard and the $\param$ value for which they have $R=1$
and turn unstable.
The last column lists the $\param$ value where the orbit has a
tangency with the boundary
and becomes unphysical for larger $\param$.} \label{tab:minimax}
\end{table}

\begin{table}[ptbh]
\centerline{
\small
\begin{tabular}{c|c|c|cc|c|c}
parent & type & sym & partner 		& partner	& $\param$
	& $\Delta\param$ \\ \hline
CC	& pd  & full  & AABB		& 		   &
0.41421356237 & 4.45e-2\\
          &    & full  & AABB  &      &  0.45873253228 & 7.34e-2\\
AABB	& pf  & T   & CABC		& CBAC		&
0.53208888624 & 2.30e-2\\
\hline
	& sn  & X   & AAABB	& ACCCC		  &
0.69532356899 & 1.87e-3\\
     &   	&	X   &	AAABB  &
& 0.69798858052	& 1.17e-2\\
AAABB	& pf  &  & CAABC		& CBAAC		&
0.70970577826 & 4.58e-3\\
\hline
	& sn  & X   & AAAABB		& AACCCC	 &
0.79431559532 & 5.55e-4\\
	&	   &	X   & AAAABB		&
& 0.79572052126	& 3.33e-3\\
AAAABB	& pf  & & AAACCB	& AAABCC	& 0.79904759065	& 1.43e-3\\
CAC=1/3	& pd  & X   & AAABAB		&	    	&
0.36810610044 & 2.84e-2\\
	  &	  &	X   & AAABAB		&
	& 0.40610662358	& 3.92e-2\\
AAABAB	& pf  & & ABAACC	& AABACC	& 0.44533792566	& 1.21e-2\\
	& sn  & full  & AAABBB	& ACCBCC	& 0.86753591308 & 3.61e-4\\
    &     & full  & AAABBB &        & 0.86789693974 & 1.41e-3 \\
AAABBB	& pf & T& BBAACC	& AABBCC	& 0.86931082693	& 5.39e-4\\
	& sn  & T    & AABABB		& ABABCC	&
0.66948391241 & 1.79e-4\\
CC	& s1/3 & T & BACCCC		& ABCCCC	&
0.37282469327	& 0 \\      
CC	& s1/3 & X & BAABCC		& ABBACC	&
0.37282469327	& 6.12e-2\\
\hline
	     & sn  & X   & AAAAABB		& AAACCCC	&
0.84879163221 & 2.31e-4\\
       	                  	  &  	&	X &	AAAAABB  &
	& 0.84943955625 & 1.35e-3\\
AAAAABB	& pf  &    & BAAAACC	& AAAABCC	& 0.85079348228	& 5.74e-4\\
	     & sn  & X   & AAAABAB		& CAACCAC	&
0.55897639858 & 2.10e-3\\
                                 	&	  &	X & AAAABAB 
		&
		& 0.56209428854 & 8.98e-3\\
AAAABAB	& pf  &   & ABAAACC		& AAABACC	&
0.57107790543	& 3.44e-3\\
        	& sn  & X   & AAAABBB  		& CAACCBC	&
0.93351335014 & 6.67e-5\\
                             	&	 &	X &	AAAABBB	    &
		& 0.93359478150 & 2.57e-4\\
AAAABBB	& pf  &   & BBAAACC		& AAABBCC	&
0.93385152715	& 9.61e-5\\
          	& sn  & X   & AABAABB		& CABACCC	&
0.61088771114 & 1.83e-4\\
	                          &	 &	X &  AABAABB	 	&
		& 0.61991127863 & 3.88e-3\\
AABAABB	& pf  &   & BAABACC		& ABAABCC	&
0.62379601274	& 1.86e-3\\
          	& sn  & X   & ACCCCCC		& BAAABCC	&
0.67586228774	& 7.77e-6\\
	                          &  &	X & 	BAAABCC	 &
		& 0.67887512963	& $-3.76$e-4\\
BAAABCC	& upf  &  & BAACCCC		& AABCCCC	& 0.67849936105	& 0\\
          	& sn  &     & AAABABB		& AABABCC	&
0.81742214822	& 1.46e-5\\
          	& sn  &     & BACCACC		& AABBACC	&
0.75730213644	& 2.35e-7\\
       \hline \end{tabular}
\vspace*{0.5cm}
}
\caption{Table of bifurcations for orbits up to period 7 in the family of
\limacon billiards. Since maximizing orbits do not bifurcate they are not
included. Note the sequences of symmetric saddle nodes
followed by a pitch fork bifurcation, see Fig.~3,4. At the end of the table
there are the first two pairs of orbits born in a saddle node bifurcation
without symmetry, see Fig.~1. For higher periods
most orbits are created in this
type of bifurcation.}\label{tab:biftab}
\end{table}

All bifurcations for periodic orbits up to period 7 are listed in
Table~\ref{tab:biftab}.
The parent orbit (which does not exist for sn) exists before and
after the bifurcation.
With the exception of the
$q$--tupling bifurcations
it changes its stability at the bifurcation,
i.e.\ it turns from elliptic to hyperbolic (for pf, pd) or vice versa
(upf, upd).
If there are two partners in a bifurcation the stable one is listed first.
In the last column of Table~\ref{tab:biftab} the width of the stable interval
is listed, where a negative number means that it extends to the left from
the listed value of $\param$. For symmetric orbits usually one of the partners
has a second interval of stability. These orbits are listed as a single partner
without a bifurcation type together with  the value $\param$ for which $R=1$.
The same families will also be parents of pitch-fork bifurcations when $R=0$.
This is listed next, so that for most symmetric orbits there are three
consecutive entries that describe the essentials of the corresponding
bifurcation tree (Fig.~3,4). For orbits with full symmetry the two 
intervals of stability
join at a point where the graph of $R(\param)$ has a tangency with 1 (Fig.~2).
In this way all stable intervals of orbits up to period 7 and
their bifurcations are listed in Table~\ref{tab:biftab}.

The relation between the winding number $p/q$ and the symbolic code for the
maximizing and minimax orbits is simple.
The code of the maximizing $p/q$ orbit is given by the word with symbols
\[
	s_i = \left\{\begin{array}{cc}
		$A$ & \mbox{ if } [ip/q] = [(i+1)p/q] \\
		$B$ & \mbox{ otherwise}
		\end{array}
\right. \;\;,
\]
where $[x]$ denotes the integer part of $x$ and $i=0,\dots,q-1$.
The corresponding minimax orbit of the same rotation number is obtained
by replacing the first and the last letter by C, see Table~\ref{tab:minimax}
for some examples.
With these rules we can predict at $\param=1$ which orbit is going to have a
particular rotation number at the integrable limit.

A typical simple saddle node bifurcation is shown in
Fig.~\ref{fig:bif_diag_AAABABB_AABABCC}.
Depending on the symmetry we observe more complicated
sequences of bifurcations, see
Fig.~\ref{fig:bif_diag_AB},
Fig.~\ref{fig:bif_diag_AAABB_ACCCC}, and
Fig.~\ref{fig:bif:diag:ACCCCCC} for examples.
These are the four basic types of (sequences of) bifurcations that occur
up to period 7, compare to the corresponding entries
in Table~\ref{tab:biftab}.


\section{Statistics of Bifurcations}

We numerically computed all stable intervals of all orbits up to period 13.
This is a complicated task that involves continuing 3697 periodic orbits.
In particular it came as a surprise that the
width of the stability intervals
for some of the orbits could not
be calculated using double precision, i.e.\ approximately 15
significant digits.
E.g.\ already at period 10 the smallest interval of stability occurs for the
saddle node pair $AAACCBCCCC$, $AAAABBCCCC$,
and is only $\Delta\param \approx 4.1 \cdot 10^{-18}$. In order to be
able to calculate such small
intervals we used the doubledouble package \cite{Bri1998}, which offers
30 significant digits. This enabled us to determine all stable intervals
of continued orbits up to period 13. Studying higher period would require
even more precision.
We now report on the results of our numerical study.

\BILD{p}
          {
          \centerline{\PSImagx{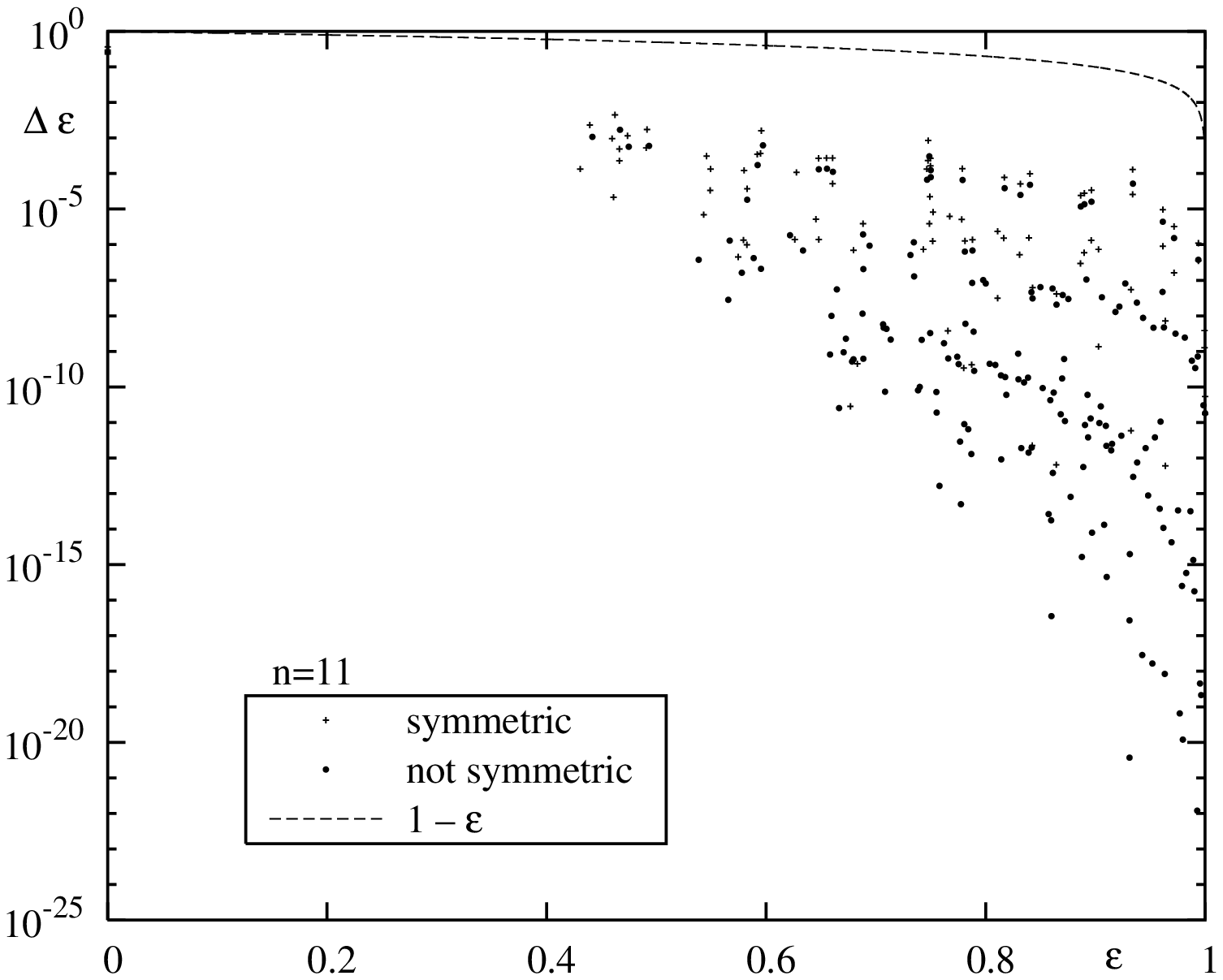}{16cm}}

          \centerline{\PSImagx{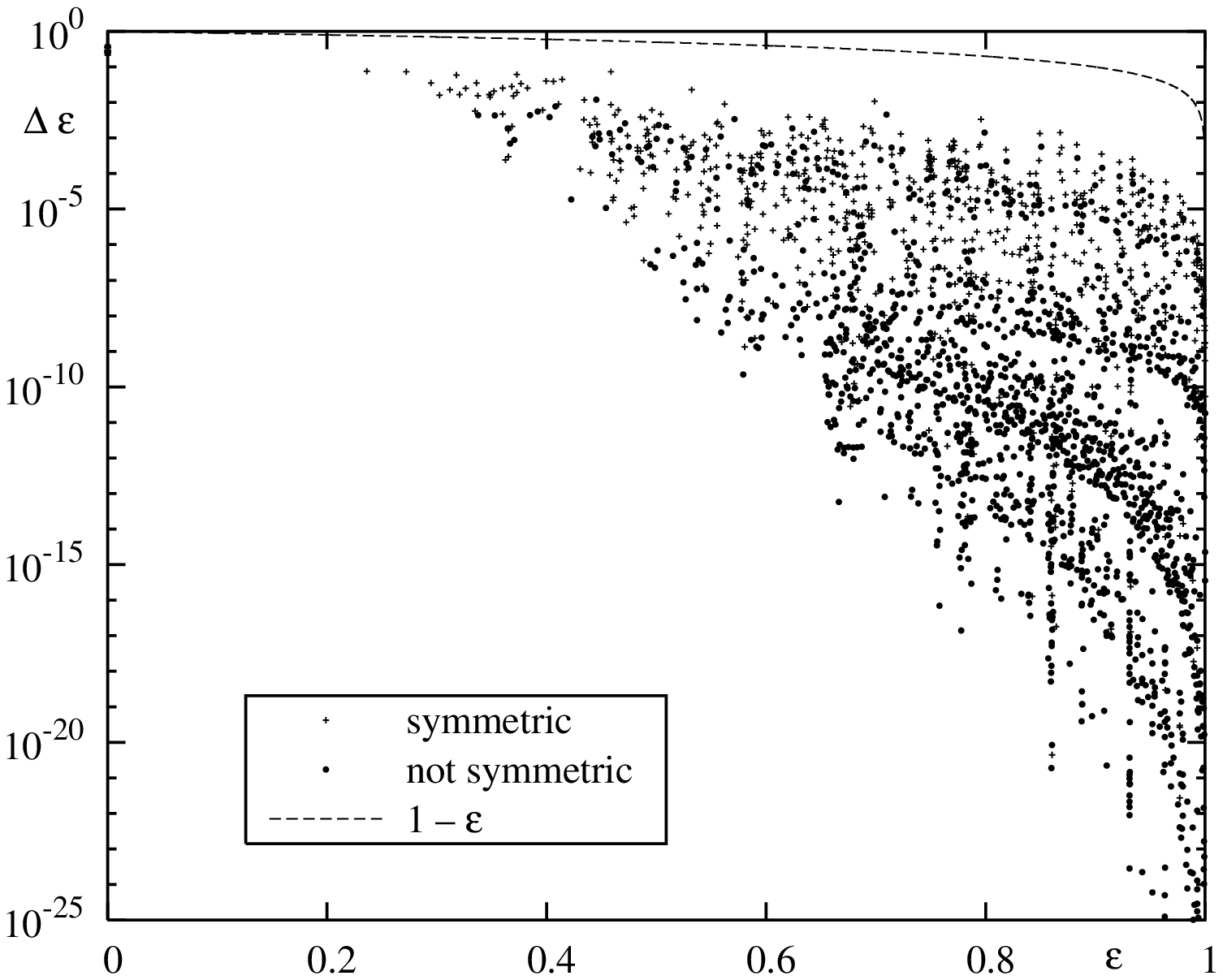}{16cm}}
          }
          {\small Width $\Delta\param$ of stable intervals as a function
of $\param$
                for orbits of period 11 and for all orbits up to period 13.
                }
          {fig:punkte}

\BILD{p}
          {
          \centerline{\PSImagx{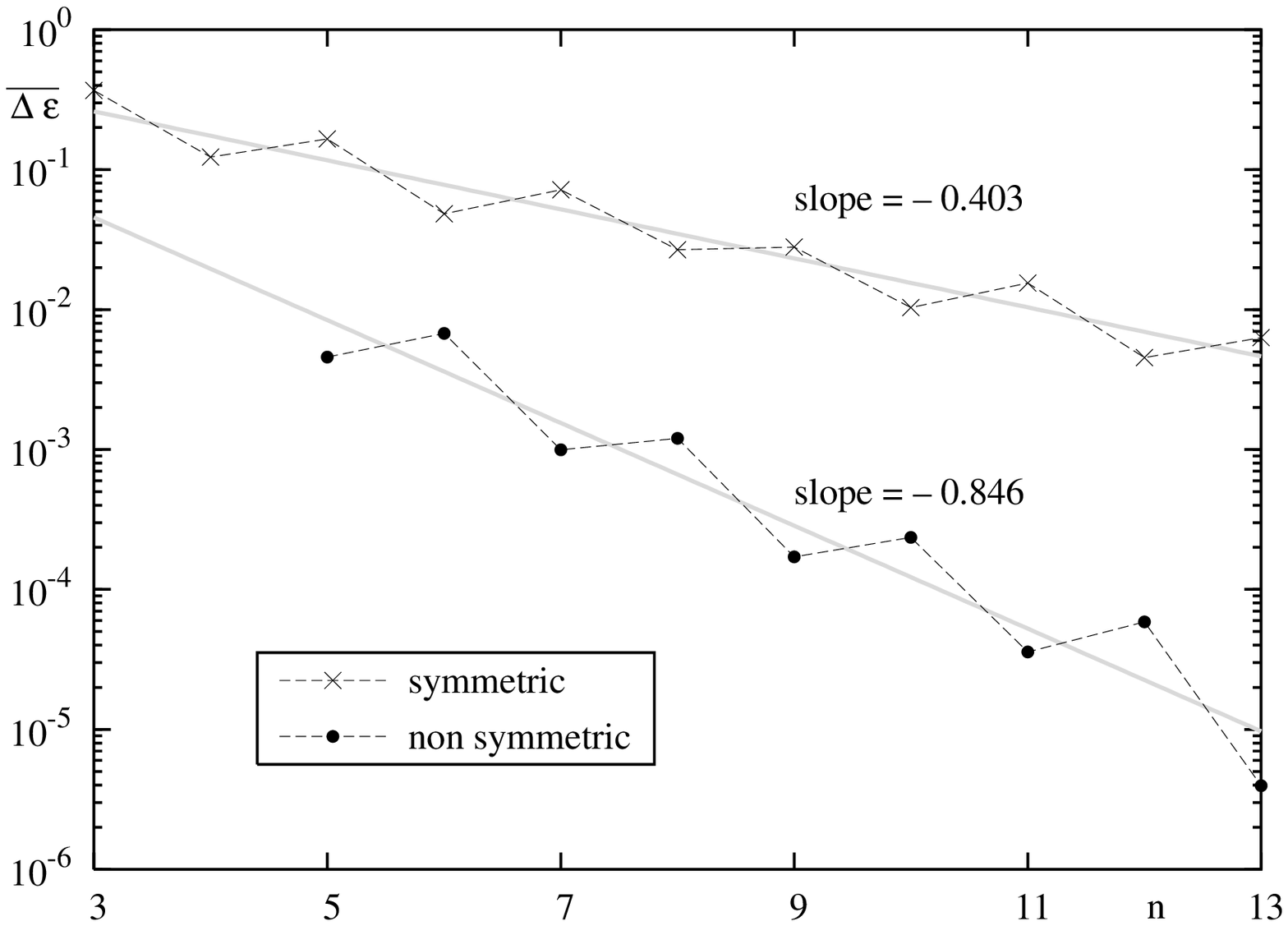}{14cm}}
          }
          {\small Averaged widths of stability intervals for symmetric and
                 non-symmetric orbits up to period 13. The slopes are
      approximately equal to the negative of the topological entropy.
                }
          {fig:avg}

\BILD{p}
          {
          \centerline{\PSImagx{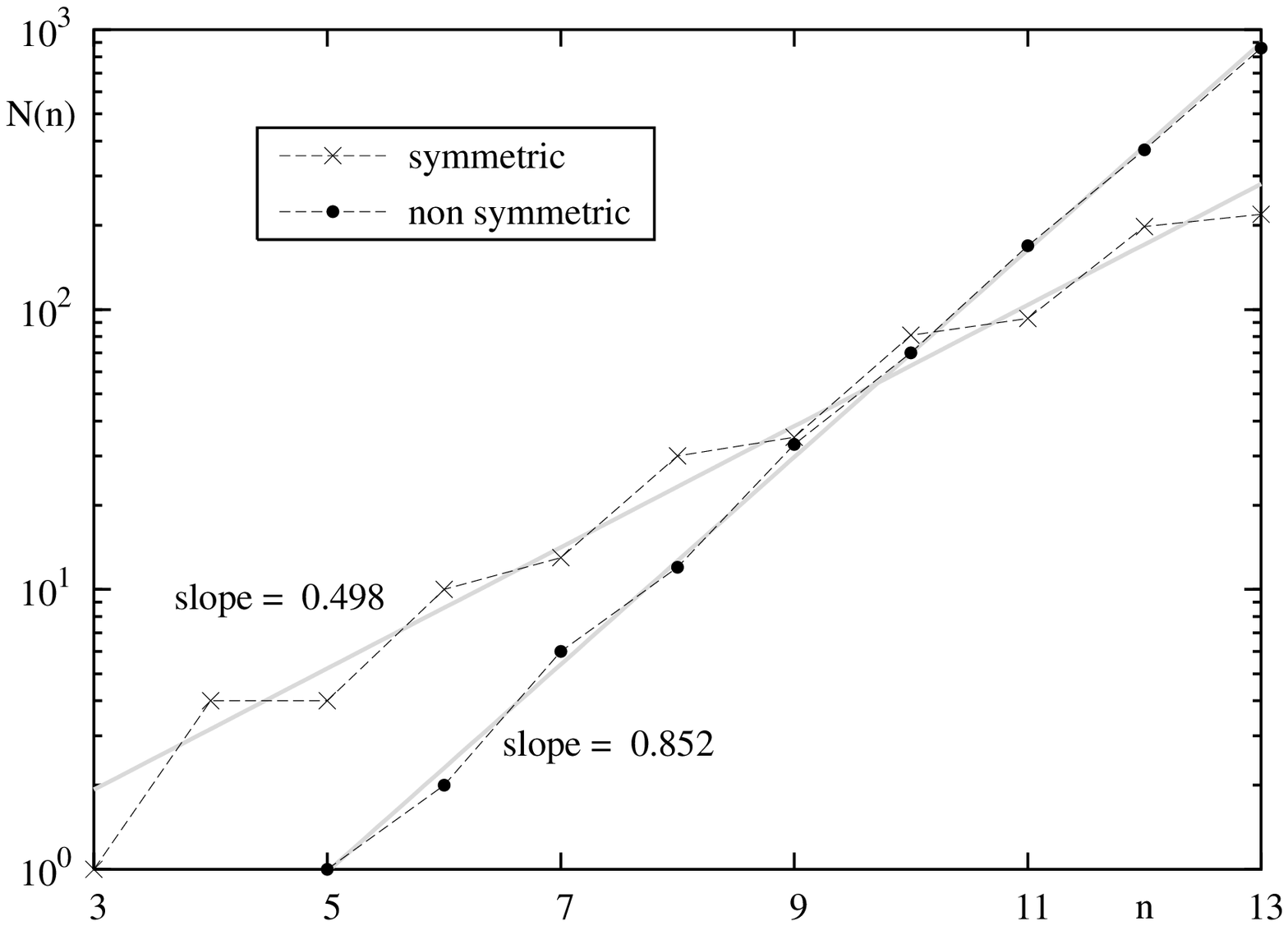}{14cm}}
          }
          {\small Number of intervals of stability as a function of 
the period up to period 13.
              The slopes approximate the topological entropy.
                }
          {fig:avg-Nn}

\BILD{p}
          {
          \centerline{\PSImagx{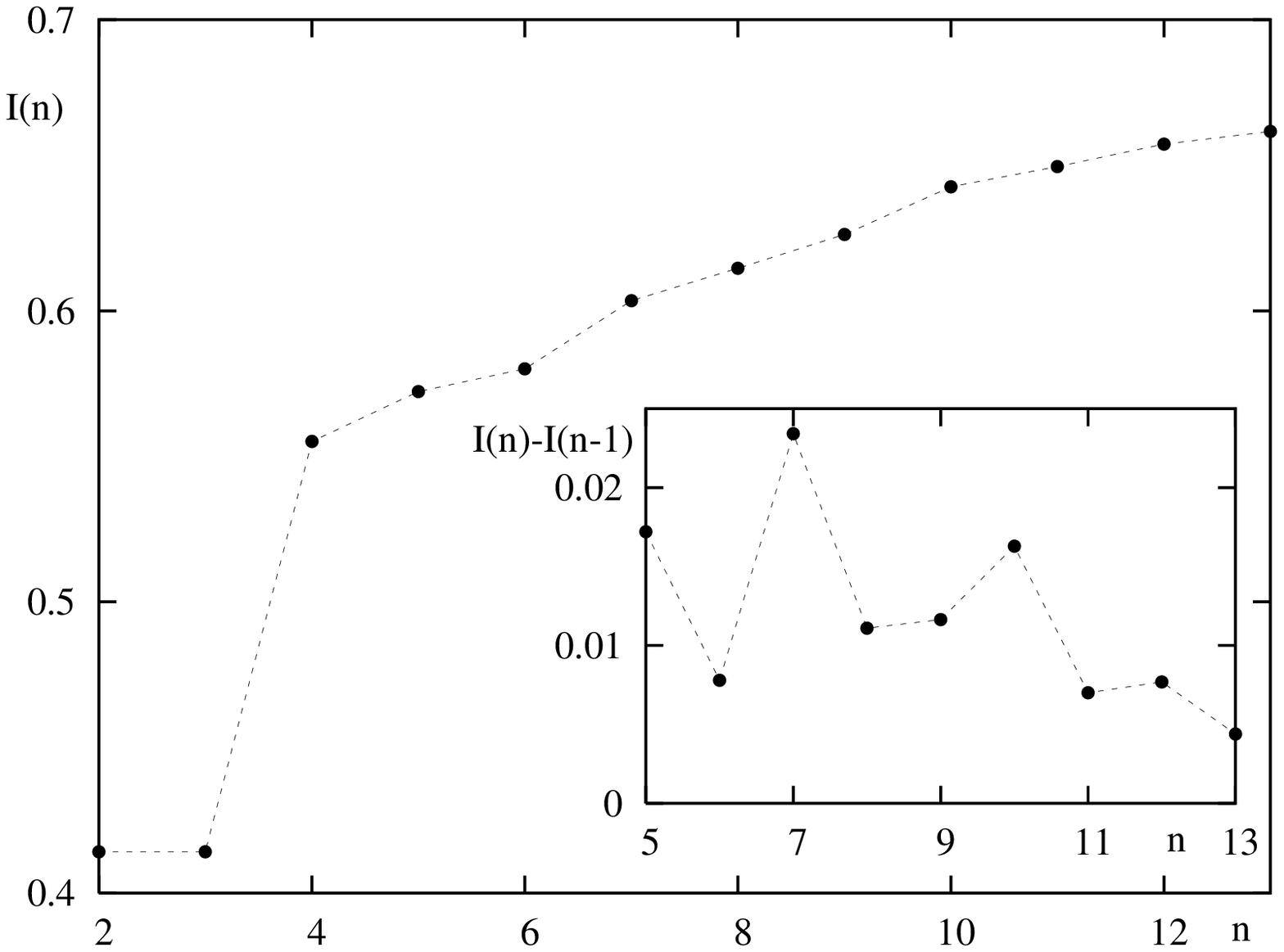}{14cm}}
          }
          {\small Total width of the union of all stable intervals up
to period 13.
                In the inset the difference between successive values is shown.
                }
          {fig:avg2}

\BILD{p}
          {
            \centerline{{\PSImagx{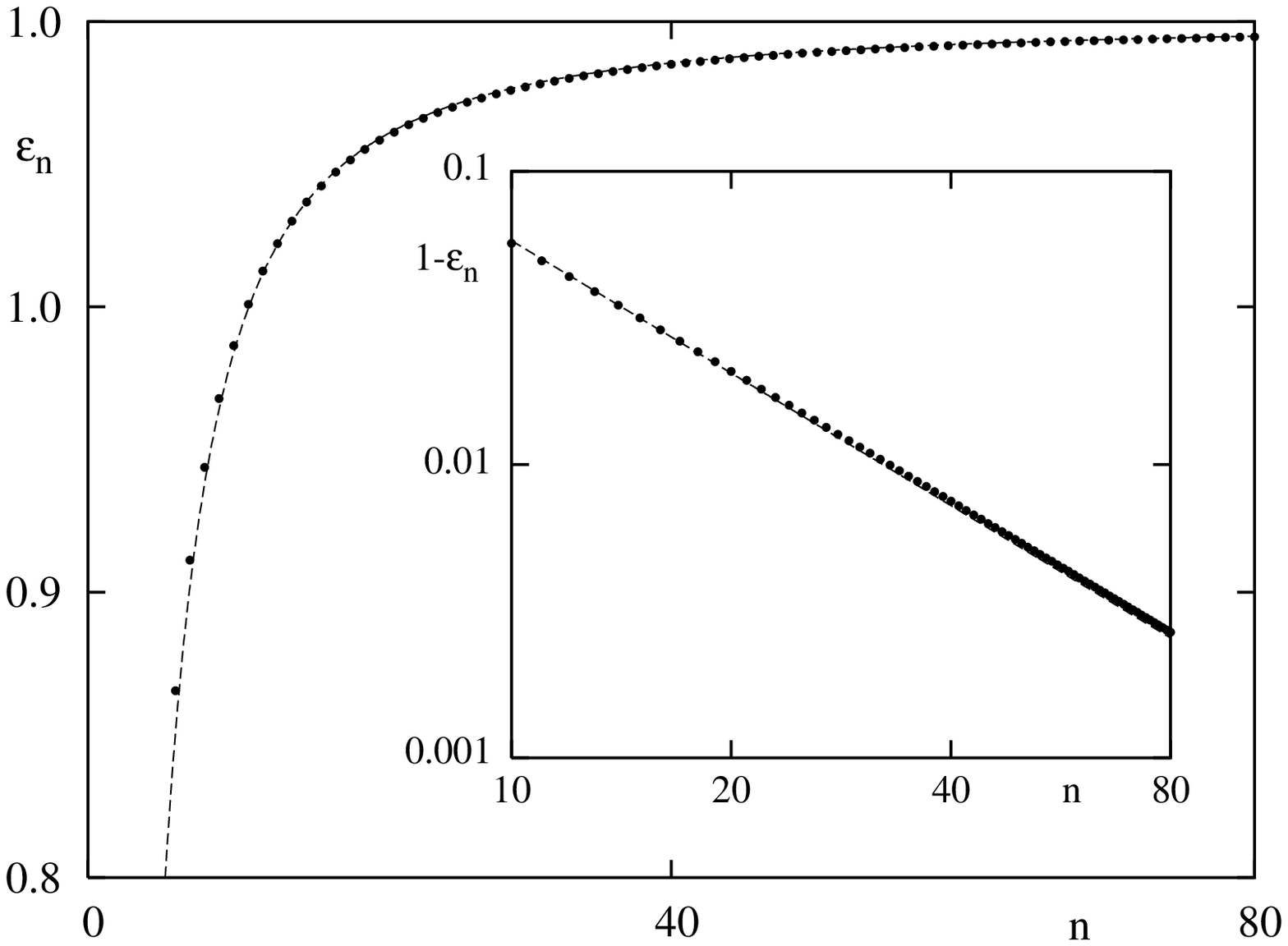}{14cm}}}
          }
          {\small The plot of the parameter of creation of $A^nBB$ orbits
                  as a function of $n$ shows that there are elliptic
                  orbits created in saddle node bifurcations arbitrarily
                   close to the cardioid billiard.
                }
          {fig:AnBB-epsilon}

Up to period 7 the results are listed in Table 2.
Fig.~\ref{fig:punkte} shows the width $\Delta\param$ of intervals of stability
as a function of $\param$
(at the left boundary of the interval) for orbits of period 11 and
for all orbits up to
period 13.
The number of points grows with the period like the number of orbits
(including cusp orbits) in the cardioid, i.e.\ exponentially with
topological entropy $\ln(1+\sqrt{2})$ \cite{BaeDul97}.
Smaller intervals occur for larger $\param$. The trivial upper bound
is $\Delta\param(\param) = 1-\param$. For fixed period a lower bound
is approximately given by a straight line. There are conspicuous patterns
that can be related to the structure of the code word.
The detailed structure is quite complicated,
and it is difficult to decide whether
for each parameter there exist elliptic orbits. Even though the number
of stable intervals grows exponentially with the period this does not
mean that they are easily visible in phase space.
In \cite{ProRob94b,LiRob95} it is referred to
an unpublished numerical study by Li and Robnik (1994),
giving that if there is a stability island for
$\varepsilon>0.555$ its relative area has to be smaller than $5\cdot 10^{-7}$.
This is in agreement with examples considered in \cite{Bae98:PhD}.

To obtain a quantitative estimate of the behaviour of width of
stability intervals
with the period we now consider the average $\overline{\Delta\param}$
of $\Delta\param$ for each period,
see Fig.~\ref{fig:avg}.
Here it is important to treat orbits with and without symmetry
separately because
orbits with symmetry tend to have larger intervals of stability.
A reason for this is that many symmetric orbits undergo symmetry
breaking pitchfork bifurcations, see
Fig.~\ref{fig:bif_diag_AAABB_ACCCC}. Therefore
they have to intersect $\TrM = 2$ with finite slope, so that
the second interval of stability is larger than the first.
There is a difference between even and odd periods. On average
non-symmetric orbits with odd period have smaller stable width than
those with even period. This is mainly due to the fact that
there are no period doubled orbits among them, which tend
to have a larger stable width because of their finite slope
in the bifurcation diagram at creation (see e.g.\ Fig.~\ref{fig:bif_diag_AB}).
The average width of the parameter interval on which the orbits are stable
follows a power law with the period.
For the non-symmetric orbits the exponent is close to the topological
entropy $0.881$. The actual growth of the number of intervals of
stability is shown in Fig.~\ref{fig:avg-Nn}.
Therefore the width of the intervals decreases at the same
rate as the number of orbits increases, such that it is not possible from
these estimates to determine the asymptotic behaviour of
the sum of all intervals.
For the symmetric orbits a approximately half the exponent is 
expected and measured.

We can conclude that the behaviour of this family of
billiards is much more complicated than one would expect on first sight.
Even arbitrarily close to the ergodic limit of the cardioid there
exist tiny elliptic islands. This can be seen by considering the
family of orbits $A^nBB$ that limits onto the whispering gallery orbit.
This family illustrates the non-uniform hyperbolicity of the
billiard map, its Liapunov exponent tends to 0 as $n$ tends to infinity,
see \cite{Bae98:PhD}.
In Fig.~\ref{fig:AnBB-epsilon} a plot of the parameter $\varepsilon_n$
where $A^nBB$ is born in a saddle node bifurcation is shown.
The data is well described by the fitted curve $\param = 1 - a n^{-3/2}$
where $a = 1.65$.
This shows that arbitrarily close to the cardioid there are stable
elliptic orbits created in saddle node bifurcations.
In Fig.~\ref{fig:punkte} this family of orbits appears approximately
as an upper bound for $\Delta \epsilon$, i.e.\ they tend to have
the largest stability intervals for given $\epsilon$.

Whether there are  parameter values for which the corresponding
\limacon billiard is
ergodic cannot be decided by a numerical experiment like the one we
performed.
The results
indicate, however,
that the set of parameter values for which the map is ergodic (if
larger than one point)
has a very complicated structure.

To our knowledge this is the first time that the scaling of
stability intervals with the period has been extensively studied.
It is well known that in hyperbolic systems there is a
scaling of the instability of periodic orbits with the
period. This expresses the fact the the phase space volume
is a constant so that growth of instability has to be exactly
compensated by growth of the number of orbits \cite{HanOzo84}.
Our results indicate that there is a similar relation for
stability intervals in parameter space for families limiting
to hyperbolic system.

\vspace{1cm}

{\bf Acknowledgements}

\vspace{0.25cm}

A.B.\ acknowledges support by the
Deutsche Forschungsgemeinschaft under contract No. DFG-Ba 1973/1-1.
H.R.D.\ was partially supported by the
Deutsche Forschungsgemeinschaft under contract No. DFG-Du 302.

\small

\end{document}